# Radio detection of Extensive Air Showers with CODALEMA

D. Ardouin[a], A. Belletoile[a], D. Charrier[a], R. Dallier[a], L. Denis[b], P. Eschstruth[c], T. Gousset[a],
F. Haddad[a], J. Lamblin[a], P. Lautridou[a], A. Lecacheux[d], D. Monnier-Ragaigne[c], O. Ravel[a]

*(a) SUBATECH, 4 rue Alfred Kastler, BP20722, F44307 Nantes cedex 3, France*
*(b) LESIA, Observatoire de Paris - Section de Meudon, 5 place Jules Janssen, F92195 Meudon cedex, France*
*(c) Observatoire de Paris - Station de radioastronomie, F18330 Nançay, France*
*(d) LAL, Université Paris-Sud, Bâtiment 200, BP 34, F91898 Orsay cedex, France*
Presenter: R. Dallier (richard.dallier@subatech.in2p3.fr), fra-lamblin-J-abs1-he15-oral

The principle and performances of the CODALEMA experimental device, set up to study the possibility of high energy cosmic rays radio detection, are presented. Radio transient signals associated to cosmic rays have been identified, for which arrival directions and shower's electric field topologies have been extracted from the antenna signals. The measured rate, about 1 event per day, corresponds to an energy threshold around $5.10^{16}$eV. These results allow to determine the perspectives offered by the present experimental design for radiodetection of UHECR at a larger scale.

## 1. Experiment and event selection

Radio emission associated with the development of Extensive Air Showers (EAS) initiated by high energy cosmic rays was investigated in the 1960's [1, 2]. A flurry of experiments provided initial informations about signals from $10^{17}$ eV cosmic rays [3], but plagued by difficulties (poor reproducibility, atmospheric effects, technical limitations) efforts almost ceased in the late 1970's to the benefit of ground particle detection [4] and fluorescence [5]. With the advent of cosmic ray research involving giant surface detectors as in the Auger experiment [6], the radio detection, with a potential 100% duty cycle and sensitive to the longitudinal development of the showers, is now reconsidered. In recent years, with the availability of new electronics, several groups have undertaken the task of reinvestigating the phenomenology of radio pulses [7, 8, 9] which are in a large extent a *terra incognita*. Using our experiment CODALEMA (COsmic ray Detection Array with Logarithmic ElectroMagnetic Antennas), located at the Nançay radio observatory, firm evidence for a radio emission counterpart of cosmic ray air showers is obtained.

Current experimental setup is shown in Fig. 1 and has been extensively described along with detection and analysis methods in Ref. [10]. It uses 11 log-periodic antennas of the type constituting the Nançay DecAMetric array (DAM) [11] and 4 particle detectors originally designed as prototype detectors for the Auger array [12].

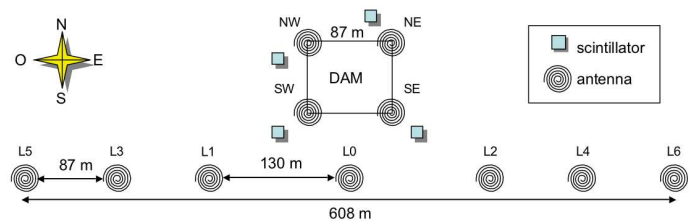

**Figure 1.** Current CODALEMA setup. The particle detectors act as a trigger with a fourfold coincidence requirement.

The antenna signals are recorded after RF signal amplification (1-200 MHz, gain 35 dB) by LeCroy digital oscilloscopes (8-bit ADC, 500 MHz sampling frequency, 10 $\mu$s recording time). To get enough sensitivity with these ADCs, the antennas are band pass filtered (24-82 MHz). Each 2.3 m$^2$ particle detector module (station) has two layers of acrylic scintillators, read out by a photomultiplier placed at the center of each sheet. The photomultipliers have copper housings and it has been firmly verified that no correlation exists between photomultiplier signal and the presence of antenna signals. The coincidence between top and bottom layers is obtained within a 60 ns time interval with a counting rate of 200 Hz per station. The whole experiment is triggered by a fourfold coincidence from those stations in a 600 ns time window. The corresponding rate is around 0.7 event per minute. Considering the active area of the particle detector array of $7.10^3$ m$^2$ and the arrival direction distribution of the shower, a value of $16.10^3$ m$^2$.sr is obtained for the acceptance, which corresponds to a trigger energy threshold of about $1.10^{15}$ eV.

For each fourfold coincidence from the particle detectors, the 11 antenna signals are recorded. Due to the relatively low energy threshold, only a small fraction of these air shower events is expected to be accompanied by significant radio signals. The recognition of the radio transients is made during an offline analysis [10]. Radio signals are first 37-70 MHz numerically filtered to detect radio transient. The maximum voltage is searched in a given time window of 2 $\mu$s width, correlated to the trigger time, and compared to a threshold based on the noise level estimation outside this window. If the threshold condition is fulfilled, the arrival time is set at the maximum voltage and the antenna is flagged. When at least 3 antennas are flagged, a triangulation procedure calculates the arrival direction of the radio wave using a plane front fit. At this level of selection, the couting rate is about one event every two hours.

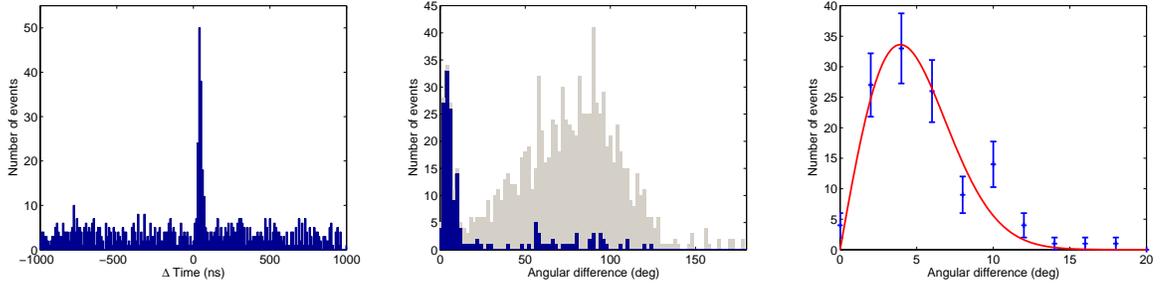

**Figure 2.** Left: Distribution of time delays between the radio plane front and the particle plane front. Middle: Distributions of angular difference between arrival directions reconstructed from antenna signals and from scintillators, without time cut (grey histogram) and with a time cut of 100 ns around the peak (blue histogram). Right: Zoom on the angular difference distribution with a time cut around the peak and fi t by the expected analytical form (red line). Errors bars are calculated as the square root of the number of events in the bin.

A stronger selection among these candidates is obtained using the arrival time distribution. It refers the radio wavefront arrival time at a particular point, as determined by antenna triangulation, to the particle front time extracted from the scintillator signals. This time difference distribution is shown in Fig. 2-left. A very sharp peak (a few tens of nanoseconds) is obtained, showing an unambiguous correlation between some radio events and the particle triggers, whereas the flat distribution corresponds to accidental radio transients which are not associated with air showers but occured in the 2 $\mu$s window where the search is conducted. Being uncorrelated to the particles, these events fill an uniform arrival time distribution. EAS events are those for which the arrival time difference between the two detector systems is within 100 ns, *i.e.* in the main peak of Fig. 2-left.

If these time-correlated events correspond to EAS, the arrival directions reconstructed from both scintillator and antenna data should be strongly correlated. Fig. 2-middle exhibits the distribution of the angle between the two reconstructed directions, without time cut (grey histogram) and with a time cut of 100 ns around the peak (blue histogram). For time-correlated events, arrival directions obtained by both particle and radio signals are the same within 15 degrees whereas the angle for an uncorrelated event is much bigger, its arrival direction given by the antennas being often close to the horizon. This is typical of events from radio interference due to human activity. The angular difference distribution of time-correlated events (Fig. 2-right) can be fit by the expected distribution, a gaussian distribution centered on zero multiplied by a sine function coming from the solid angle factor. The standard deviation of the corresponding gaussian is about 4 degrees. This value includes the reconstruction accuracy of both detectors.

After this sorting, confirmed EAS radio event counting rate falls down to 1 event per day, which corresponds to an energy threshold of about $5.10^{16}$ eV for the antenna array with an assumed acceptance of $16.10^3$ $m^2$.sr. At the end of these analysis procedures, physical characteristics of the radio EAS events can be extracted.

## 2. EAS electric field distribution

Due to the limited extent of the North-South axis of the array, the electric field distribution have mainly been extracted from an analysis along the East-West axis. Antenna responses were cross-calibrated and gains adjusted within a few %. Distributions for 4 illustrative radio EAS events are shown on the Fig. 3 (with a gaussian fit to guide the eye). A fortuitous event (triangles) and the threshold level of our setup illustrating the amount of useful signal received (circles) are also plotted. The fortuitous event was identified as resulting from an anthropic polluting source and rejected from EAS candidate status.

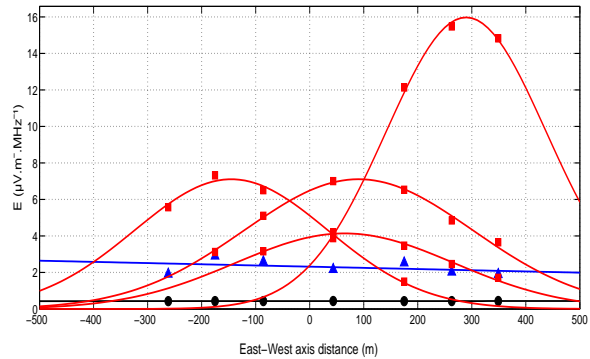

**Figure 3.** Profile of the maximum voltage (corrected from the cable attenuation and the amplifier gains) recorded on the antennas in the East-West direction for 4 EAS events (squares, gaussian fit: red/full line), an anthropic transient (triangles, fit: blue/dashed line) and detection thresholds (circles, fit: black/dash-dotted line).

Topologies are clearly different between EAS and anthropic events. The anthropic event presents an electric field topology with a nearly linear amplitude which is not expected for an EAS candidate falling in the vicinity of the array. EAS events show highly variable field amplitudes depending on the position on the E-W axis. The width of the distributions observed on an event by event basis indicates probably the dependence on both shower energy and zenithal angle. The projected core locations have been estimated only for events falling inside the surface delimited by the extremities of our antenna array (highly limited by the North-South extent) [13]. The margin between electric field topologies depending on its origin (EAS or anthropic) could constitute one decisive criterion of selection as it comes from the antenna array only and not from a comparison to the particle detectors. In other words, it means that a radiodetection of cosmic ray experiment should be able to discriminate EAS events by itself.

## 3. Conclusions

Electric field transients generated by extensive air showers have been measured with CODALEMA. The current effective counting rate of 1 event/day leads to an energy threshold around $5.10^{16}$ eV. Electric field spread related to energy and core location of an EAS can be determined on an event by event basis. It is now possible to discriminate an EAS event from a fortuitous one using only antennas and no particle detector. This is one further step towards a stand-alone system that could be deployed over a large area. Nevertheless, more data and technical upgradings are needed to examine the contribution that radio detection could bring in determining the energy and the nature of the cosmic rays. Improvements are in progress: setting up of additional scintillators will make possible an independant determination of the shower energy and core position; extension of the N-S antenna line will enable to better sample the radio signal spread; increase of the ADC dynamics using 12-bit encoding will allow to record the full 1 MHz - 100 MHz frequency band, and shower parameters could then be inferred from the full signal shape [10]. In a subsequent upgrade, it is planned to install autonomous dipoles equipped with active front-end electronics, self-triggered and self-time-tagged. This is part of current investigations on the feasibility of adding radio detection techniques to an existing surface detector such as the Pierre Auger Observatory. In the future, we except that the radio signals should provide complementary information about the longitudinal development of the shower, as well as the ability to lower the energy threshold.

**References**

[1] G.A. Askar'yan, Soviet Physics, J.E.T.P., 14, (2) 441 (1962)
[2] T.C. Weekes, Proc. of the first int. workshop on "Radiodetection of high Energy Particles", Los Angeles, November 16-18, 2000, AIP Conference Proceedings Vol 579 (2001) p. 3-13,
[3] H.R. Allan, in: Progress in elementary particle and cosmic ray physics, ed. by J.G. Wilson and S.A. Wouthuysen (North Holland, 1971), p. 169.
[4] N. Hayashida et al., Phys. Rev. Lett. 73, 3491 (1994); M. Takeda et al., Phys. Rev. Lett. 81, 1163 (1998).
[5] D.J. Bird et al., Phys. Rev. Lett. 71, 3401 (1993); Astrophys. J. 441, 144 (1995).
[6] Auger Collaboration, Pierre Auger Project Design Report (2nd Edition, November 1996, revised March 1997), available from http://www.auger.org; J.W. Cronin, Rev. Mod. Phys. 71, S165 (1999).
[7] K. Green, J.L. Rosner, D.A. Suprun, J.F. Wilkerson, Nucl. Instrum. Meth. A498, 256 (2003).
[8] H. Falcke, W. D. Apel, A. F. Badea, et al (LOPES Collaboration), Nature, May 19, 2005.
[9] I. Kravchenko et al., astro-ph/0306408.
[10] D. Ardouin, A. Bellétoile, D. Charrier, R. Dallier, L. Denis, P. Eschstruth, T. Gousset, F. Haddad, J. Lamblin, P. Lautridou, A. Lecacheux, D. Monnier-Ragaigne, A. Rahmani, O. Ravel, to appear in Nucl. Inst. and Meth. - A (2005), astro-ph/0504297.
[11] http://www.obs-nancay.fr/
[12] M. Boratav, J.W. Cronin, B. Dudelzak, P. Eschstruth, P. Roy, V. Sahakian and Z. Strachman, The AUGER Project: First Results from the Orsay Prototype Station, Proceedings of the 24th ICRC, Rome, 954 (1995).
[13] D. Ardouin, A. Bellétoile, D. Charrier, R. Dallier, L. Denis, P. Eschstruth, T. Gousset, F. Haddad, J. Lamblin, P. Lautridou, A. Lecacheux, D. Monnier-Ragaigne, O. Ravel, in Proceedings of the XXXIXth Rencontres de Moriond "Very High Energy Phenomena in the Universe", La Thuile, Italy (2005), astro-ph/0505442.